\documentclass[aps,prd,preprint,showpacs,english]{revtex4}
\usepackage[T1]{fontenc}
\usepackage{graphicx}
\usepackage{babel}
\usepackage{amsmath}

\usepackage{epsfig}

\usepackage{ mathrsfs }

\begin{document}

\title{A $6$D standing-wave Braneworld with normal matter as source.}
\author{L. J. S. Sousa}
\email{luisjose@fisica.ufc.br}
\affiliation{Instituto Federal de Educa\c{c}\~{a}o, Ci\^{e}ncia e Tecnologia do Cear\'a (IFCE), Campus Canind\'{e}, 62700-000 Canind\'{e}-Cear\'{a}-Brazil}
\author{W. T. Cruz}
\email{wilami@fisica.ufc.br}
\affiliation{Instituto Federal de Educa\c{c}\~{a}o, Ci\^{e}ncia e Tecnologia do Cear\'{a} (IFCE), Campus Juazeiro do Norte, 63040-000 Juazeiro do Norte-Cear\'{a}-Brazil}

\author{C. A. S. Almeida}
\email{carlos@fisica.ufc.br}
\affiliation{Universidade Federal do Cear\'{a}, Departamento de F\'{i}sica - Campus do Pici, \\ C.P. 6030, 60455-760, Fortaleza-Cear\'{a}-Brazil}

\begin{abstract}
A six dimensional  standing wave braneworld model has been constructed. It consists in an anisotropic 4-brane generated by standing gravitational waves whose source is normal matter. In this model,  the compact (on-brane) dimension is assumed to be sufficiently small in order to describe our universe (hybrid compactification). The bulk geometry  is non-static, unlike most of the braneworld models in the literature. The principal feature of this model is the fact that the source is not a phantom like scalar field, as the original standing-wave model that was proposed in five dimensions and its six dimensional extension recently proposed in the literature. Here, it was obtained a solution in the presence of normal matter what assures that the model is stable. Also, our model is the first standing wave brane model in the literature which can be applied successfully to the hierarchy problem. Additionally, we have shown that the zero-mode for the scalar and fermionic fields are localized around the brane. Particularly for the scalar field we show that it is localized on the brane, regardless the warp factor is decreasing or increasing. This is in contrast to the case of the local string-like defect where the scalar field is localized for a decreasing warp factor only.
\end{abstract}


\maketitle

\section{Introduction}
\label{sec:introduction}

The so called braneworld models assume the idea that our universe is a membrane, or brane, embedded in a higher-dimensional space-time. The success of this idea between the 	
physicists can be explained basically because these models have brought a solution for some insoluble problems in the Standard Model (SM) physics, as the hierarchy problem. There are many theories that carry this basic idea, but the main theories in this context are the one first proposed by Arkani-Hamed, Dimopoulos and Dvali \cite{I.Antoniadis1998, N.Arkani-Hamed1999, N.Arkani-Hamed1999a} and the so-called, Randall-Sundrum (RS) model \cite{Randall1999, Randall1999a}.

In  these models, it is assumed a priori that all the matter fields are restricted to propagate only in the brane. The gravitational field is the only one which is free to propagate in all the bulk. However, some authors  have argued that this assumption is not so obvious, and it is necessary to look for alternative theoretical mechanisms of field localization in such models \cite{Oda:2000zc,Oda2000a}. Accordingly, before to study the cosmology of a braneworld model, it is convenient to analyze its capability to localize fields. Therefore, for a braneworld model to be indicated as a potential candidate of our universe it is necessary to be able to localize the Standard Model fields.

The Randall-Sundrum model was generalized to six dimensions by several interesting works \cite{Oda:2000zc,Oda2000a,Gregory:1999gv,Chen:2000at,Cohen:1999ia,Olasagasti:2000gx,Gherghetta:2000qi,Ponton:2000gi,
Giovannini:2001hh,Tinyakov:2001jt,Kanno:2004nr,Vinet:2004bk,Cline:2003ak,Papantonopoulos:2007fk,Navarro:2003vw,Navarro:2004di,
Papantonopoulos:2005ma,deCarlos:2003nq,Gogberashvili:2001jm,Koley2007,Torrealba,Silva:2011yk,Luis2012, Merab2012}.
A high number of the works in six dimensions refer to the scenarios where the brane has cylindrical symmetry,  the so-called string-like braneworlds, which is associated to topological defects. Some of these six dimensional models are classified as global string \cite{Oda:2000zc,Gregory:1999gv,Olasagasti:2000gx},
the local string \cite{Gherghetta:2000qi,Ponton:2000gi,Giovannini:2001hh}, thick string \cite{Kanno:2004nr,Vinet:2004bk,Papantonopoulos:2007fk,Navarro:2003vw,Navarro:2004di,Papantonopoulos:2005ma} and supersymmetric cigar-universe \cite{deCarlos:2003nq}. Also, the work proposed here is a generalization of the  RS model for six dimensional space-time. However we treat the so called standing wave brane model which will be discussed later.

On the other hand, studies of field localization are  very  common in the literature in 5D \cite{Merab2011a,Merab2011b, Kehagias2001,Cruz2009,W.T.Cruz2011,Tahim2009} and 6D braneworld \cite{Gherghetta:2000qi,Oda:2000zc,Oda2000a,Giovannini:2001hh,Torrealba,Silva:2011yk,Luis2012}. In general, we find strategies of localization for all the Standard Model fields but the way that this localization is possible varies in different works. In some of them, the localization is possible by means of gravitational interactions only \cite{Oda:2000zc,Oda2000a}. In other works, it is necessary to consider the existence of auxiliary fields, like the dilaton \cite{Cruz2009,W.T.Cruz2011}. As far as we know, there is not in the literature a purely analytical geometry that localizes all the SM fields by means of the gravitational field interaction only. Hence, looking for a model which presents the features of to be analytical and be able to localize all the Standard Model fields is, in our point of view, an appropriated reason to study  localization of fields in different braneworld models.

The search for such a model has motivated the appearance  of some braneworld scenarios with non-standard transverse manifold. Randjbar-Daemi and
Shaposhnikov obtained trapped massless gravitational modes and chiral fermions as well, in a model that they called  a Ricci-flat space or a homogeneous space \cite{RandjbarDaemi:2000ft}. Kehagias proposed an interesting model which drains the vacuum energy, through a conical tear-drop like space which forms a transverse space with a conical singularity. In this way, it was possible to explain the small value of the cosmological constant \cite{Kehagias:2004fb}. Another non-trivial geometry was proposed by Gogberashvili \textit{et al}. They have found  three-generation for fermions on a $3$-brane whose transverse space has the shape of an apple \cite{Gogberashvili:2007gg}. It is still possible to cite other examples of space used like the torus \cite{Duan:2006es}, a space-time geometry with football-shape  \cite{Garriga:2004tq} and  smooth versions of the conifold, classified as  resolved conifold \cite{VazquezPoritz:2001zt} and deformed conifold \cite{Firouzjahi:2005qs, Noguchi:2005ws, Brummer:2005sh}.

The standing wave braneworld  was first proposed in five dimensions by  Gogberashvili and Singleton \cite{GOGBERASHVILI1}. This is a completely anisotropic braneworld model whose source is a phantom-like scalar (a scalar field with a wrong sign in front of the kinetic term in the Lagrangian). To avoid the problem with instability, normally presented in theory with phantom scalar, the model is embedded in a 5D Weyl geometry in such a way that the phantom-like scalar may be associated  with the Weyl scalar \cite{Merab2011a,Merab2011b,Gogberashvili:2010yf}, which  is stable. About the Weyl scalar, we may point out their presence also in other braneworld scenarios like the pure-gravity, which is an extension of the RS model. In the context of field localization in the standing wave approach, it was possible to localize several fields in five dimensions, albeit the right-handed fermions were not localized neither in increasing nor decreasing warp geometry \cite{Merab2011a,Merab2011b}. It is worthwhile to mention that the models generated by  phantom-like scalar are relevant  phenomenologically since this exotic source is useful in different scenarios like cosmology \cite{Koley2007}, where the phantom scalar is used to explain dark energy theories and the accelerated expansion of the universe \cite{Caldwell et al}. An extension for 6D of the standing wave 5D model with a phantom like scalar was first proposed by some authors of this article \cite{Luis2012c}. Additionally, the study of massive modes was not addressed in this model in five dimensions or even its six dimensional version.

In this article, we do not specify \textit{a priori} the  source or the \textit{stuff} from which the brane is done. We consider a general matter source and look for a standing wave solution. In contrast to the work of Gogberashvili and collaborators in the 5D model \cite{Merab2011a}, in which the source is a phantom-like scalar field, here we have obtained standing gravitational waves solutions of Einstein equations in the presence of \textit{normal matter} (we are using the classification for different types of matter given by M. Visser \cite{VISSER}). Since it is done by normal matter, the model constructed here is stable.  Our model with normal matter as a source is a first six dimensional one, but quite recently Midodashvili \textit{et al.} \cite{Midodashvili2012} constructed a 5D standing wave braneworld model with a real field as a source.

The model built here consists of a 6D braneworld with an anisotropic 4-brane, where the small, compact dimension belongs to the brane. The bulk is completely anisotropic, unless for some points called the $AdS$ islands \cite{Merab2011a,Merab2011b}. Its dynamics, as in the case of the works of Gogberashvili and collaborators and its extensions, represents a special feature in the sense that both metric and source are time dependent. We present two types of solutions: one with an isotropic cosmological constant where the source despite the fact that all its components are positive do not satisfy the dominant energy condition - DEC. This source may be classified as a \textit{not normal matter} \cite{VISSER}. In the other case, we make use of a recently proposed approach that suggests an extension for the Randall-Sundrum model to higher dimensions in the presence of an anisotropic cosmological constant \cite{ARCHER}. In this case, we find solutions in the presence of normal matter.

We have obtained analytical solution for the warp factor, which corresponds to a thin brane, for both decreasing and increasing warp factor. The bulk is smooth everywhere  and converges asymptotically to an $AdS_{6}$ manifold. We have considered a minimally coupled scalar field, and we have shown that it is localized in this model.  Here, we have obtained results that are more general that those encountered  for the string-like defect and the 5D and 6D versions of the standing wave approach.  Indeed, here  the scalar field is trapped for both decreasing and increasing  warp factor whereas in the string-like is a localized mode for a decreasing warp factor only. Also in 5D and 6D versions of the standing wave models the scalar field is localized for increasing warp factor only.

Furthermore, our six dimensional standing wave braneworld with physical source is an interesting scenario in order localize fermions fields. Indeed, we show that right-handed fermions can be localized in this brane.

We organize this work as follows: in section (\ref{model}) the model is described and the Einstein equations are solved in order to obtain the general expressions for the  source and the function that characterizes the anisotropy. In section (\ref{solution}), we have found the standing waves solutions, and we have discussed its mainly features. We still show that the energy-momentum components are all positives. In the case of an anisotropic cosmological constant they obey all the energy conditions which characterizes a normal matter source. The localization of the zero mode of scalar and fermionic fields have been done in the sections (\ref{scalar}) and (\ref{fermion}), respectively. Some remarks and conclusions are outlined in section (\ref{conclusions}).

\section{The model}
\label{model}

Our intent is to derive a standing wave solution of the Einstein equations, by considering normal matter as source. So, we consider the standard Einstein-Hilbert action in six dimensional space-time added by a matter field action which may be time dependent. Namely,
\begin{equation}\label{action}
S = \frac{1}{2\kappa_{6}^{2}} \int d^{6}x \sqrt{- g}\Big[(R - 2\Lambda_{6}) + L_{m}\Big],
\end{equation}
where $\kappa _{6}$ is the six dimensional gravitational constant, $\Lambda_{6}$ is the bulk cosmological constant and $L_{m}$ is any matter field Lagrangian.

From the action (\ref{action}) we derive the Einstein equations
\begin{equation} \label{Einstein}
R_{MN}-\frac{1}{2} g_{MN}R = -\Lambda g_{MN} + \kappa_{6}^{2}T_{MN},
\end{equation}
where M, N,... denote D-dimensional space-time indices and the $T_{MN}$ is the energy-momentum tensor defined as
\begin{equation}
T_{MN} = - \frac{2}{\sqrt{- g}} \frac{\delta}{\delta g^{MN}} \int d^{6}x \sqrt{- g} L_{m}.
\end{equation}

The general ansatz for the metric considered in this  work is given as follows
\begin{equation}
\label{gen-metric}
ds^{2}= e^{A}\left( - dt^{2} + e^{u}dx^{2} + e^{u}dy^{2} + e^{-3u}dz^{2} \right) + dr^{2} + R_{0} ^{2}e^{B + u}d\theta^{2},
\end{equation}
where the functions $A(r)$ and $B(r)$ depend only on $r$ and the function $u$ depends on $r$ and $t$ variables. For this metric ansatz, $(\ref{gen-metric})$,  the Einstein equations (\ref{Einstein}) may be rewritten as
\begin{eqnarray}
\label{Einstein-xx1}
\lefteqn{G_{xx} = G_{yy} = \left(\frac{1}{4} e^{A + u}\right)}\nonumber\\
& &\left( 6A^{'2} + B^{'2} + 3A^{'}B^{'}+ 6A^{''} + 2B^{''} + 6(u^{'2} - e^{-A}\dot{u}^{2}) + 2 e^{-A}\ddot{u} - 5A^{'}u^{'} -2 u^{''}\right) \nonumber\\ & & = \kappa_{6}^{2}T_{xx} - e^{A + u} \Lambda_{6},
\end{eqnarray}
\begin{eqnarray}
\label{Einstein-zz1}
\lefteqn{ G_{zz} = \left(\frac{1}{4} e^{A -3 u}\right)}\nonumber\\
& &\left( 6A^{'2} + B^{'2} + 3A^{'}B^{'}+ 6A^{''} + 2B^{''} + 6(u^{'2} - e^{-A}\dot{u}^{2}) -6 e^{-A}\ddot{u} + (11A^{'} + 4 B^{'})u^{'} + 6 u^{''}\right) \nonumber\\ & & =  \kappa_{6}^{2} T_{zz} - e^{A -3 u} \Lambda_{6},
\end{eqnarray}
\begin{eqnarray}
\label{Einstein-tt1}
\lefteqn{ G_{tt} = \left(\frac{1}{4} e^{A}\right)}\nonumber\\
& &\left( -6A^{'2} - B^{'2} - 3A^{'}B^{'} - 6A^{''} - 2B^{''} - 6(u^{'2} + e^{-A}\dot{u}^{2})  + (A^{'} -  B^{'})u^{'} \right) \nonumber\\ & & =  \kappa_{6}^{2} T_{tt} + e^{A} \Lambda_{6},
\end{eqnarray}
\begin{equation}
 \label{Einstein-rt1}
G_{rt} = \frac{1}{4}\dot{u}(A^{'} - B^{'} - 12 u^{'})= \kappa_{6} ^{2}T_{r t},
\end{equation}
\begin{eqnarray}
\label{Einstein-rr1}
\lefteqn{ G_{rr} = \left(\frac{1}{4} \right)}\nonumber\\
& &\left( 6A^{'2}  + 4A^{'}B^{'}  - 6(u^{'2} + e^{-A}\dot{u}^{2})  + (A^{'} -  B^{'})u^{'} \right) \nonumber\\ & & =  \kappa_{6}^{2} T_{r r} - \Lambda_{6},
\end{eqnarray}
and
\begin{eqnarray}
\label{Einstein-teta1}
\lefteqn{ G_{\theta \theta} = \left(\frac{1}{4}R_{0} ^{2} e^{B + u}\right)}\nonumber\\
& &\left( 10 A^{'2} + 8A^{''}  + 6(u^{'2} - e^{-A}\dot{u}^{2}) + 2 e^{-A}\ddot{u} -5A^{'}u^{'} - 2 u^{''}\right) \nonumber\\ & & = \kappa_{6}^{2}T_{\theta \theta} - R_{0} ^{2} e^{B + u} \Lambda_{6}.
\end{eqnarray}

The case $A = B = 2 a r$ was treated in a previous work \cite{Luis2012c}. As a matter of fact, in this case, it is possible to find a standing gravitational wave solution in the presence of a phantom-like scalar field, similar to the one first found in five dimensions. Another 6D standing wave braneworld has been recently proposed \cite{Midodashvili2012}. However, in that case the metric is quite different from the one considered here as given by Eq. (\ref{gen-metric}).  In this last model the spatial metric components $x, y, z$ are all multiplied by the same factor $e^{2 a r + u}$, while the compact extra dimension is multiplied by $e^{2 a r - 3 u}$. The solution, in this case, is similar to the one found in Ref. \cite{Luis2012c} and the source is still a phantom-like scalar field. The two models are still similar in the results of field localization.

As was commented above, the two 6D standing wave braneworld models recently proposed in the literature have shown interesting results in field localization, but both are generated by exotic matter, a phantom like scalar. Here, we are interested in studying the possibility to have a standing wave braneworld generated by normal matter whereas maintains the efficiency in localizing fields. So we will consider the case where $A \neq B$, $A(r) = 2 c r$, and $B(r) = c_{1} r$. In this case, the set of equations (\ref{Einstein-xx1} - \ref{Einstein-teta1}) will be simplified to
\begin{eqnarray}
\label{Einstein-xx5}
\lefteqn{ \left( \frac{1}{4} \right)}\nonumber\\
& & \left( 24 c^{2} + c_{1} ^{2} + 6 c c_{1}  + 6(u^{'2} - e^{-2c r}\dot{u}^{2}) + 2 e^{-2 c r}\ddot{u} - 10 c u^{'} -2 u^{''} \right) \nonumber\\
& & = \kappa_{6}^{2}T^{x}_{ x} -  \Lambda_{6},
\end{eqnarray}
\begin{eqnarray}
\label{Einstein-zz1}
\lefteqn{ \left(\frac{1}{4} \right)}\nonumber\\
& &\left( 24 c^{2} + c_{1} ^{2} + 6 c c_{1} + 6(u^{'2} - e^{-2 c r}\dot{u}^{2}) -6 e^{- 2 c r}\ddot{u} + (22 c + 4 c_{1})u^{'} + 6 u^{''}\right) \nonumber\\ & & =  \kappa_{6}^{2}T^{z}_{ z} -  \Lambda_{6},
\end{eqnarray}
\begin{eqnarray}
\label{Einstein-tt1}
\lefteqn{ - \left(\frac{1}{4} \right)}\nonumber\\
& &\left( -24 c^{2} - c_{1} ^{2} - 6 c c_{1} - 6(u^{'2} + e^{-2 c r}\dot{u}^{2})  + (2 c -  c_{1})u^{'} \right) \nonumber\\ & & =  \kappa_{6}^{2}T^{t}_{ t} -  \Lambda_{6},
\end{eqnarray}
\begin{equation}
 \label{Einstein-rt1}
 \frac{1}{4}\dot{u}(2 c - c_{1} - 12 u^{'})= T_{r t},
\end{equation}
\begin{eqnarray}
\label{Einstein-rr1}
\lefteqn{  \left(\frac{1}{4} \right)}\nonumber\\
& &\left( 24 c^{2}  + 8 c c_{1}  - 6(u^{'2} + e^{-2 c r}\dot{u}^{2})  + (2 c -  c_{1})u^{'} \right) \nonumber\\ & & =  \kappa_{6}^{2}T^{r}_{ r} -  \Lambda_{6},
\end{eqnarray}
and
\begin{eqnarray}
\label{Einstein-teta1}
\lefteqn{  \left( \frac{1}{4} \right)}\nonumber\\
& &\left( 40 c^{2} +  6(u^{'2} - e^{-2 c r}\dot{u}^{2}) + 2 e^{-2 c r}\ddot{u} -10 c u^{'} - 2 u^{''} \right) \nonumber\\ & & = \kappa_{6}^{2}T^{\theta} _{ \theta} -  \Lambda_{6}.
\end{eqnarray}

In order to have standing wave solution we will choose
\begin{equation} \label{Einstein-simplif1}
- e^{-2 c r}\ddot{u} + \frac{1}{6}(22 c + 4 c_{1})u^{'} + u^{''}=0.
\end{equation}

In this case, the energy-momentum components have to satisfy the relations
\begin{equation} \label{EM-phantomlike-xx1}
\kappa_{6}^{2} T^{x} _{x} =  \kappa_{6}^{2}T^{y} _{y} = \frac{1}{4}\left( 6(u'^{2} - e^{-2 c r}\dot{u}^{2}) - \frac{4}{3}(2 c - c_{1})u' + 6 c c_{1}\right),
\end{equation}

\begin{equation} \label{EM-phantomlike-zz1}
\kappa_{6}^{2} T^{z} _{z}  = \frac{1}{4}  \left(6(u'^{2} - e^{-2 c r}\dot{u}^{2})) + 6 c c_{1}\right),
\end{equation}
\begin{equation} \label{EM-phantomlike-tt1}
\kappa_{6}^{2} T^{t} _{t}  = - \frac{1}{4}   \left(-6(u'^{2} + e^{-2 c r}\dot{u}^{2}) + (2 c - c_{1})u' -6 c c_{1} \right),
\end{equation}

\begin{equation} \label{EM-phantomlike-rr1}
\kappa_{6}^{2} T^{r} _{r}  =  \frac{1}{4} \left( -6(u'^{2} + e^{- 2 c r}\dot{u}^{2}) + (2 c - c_{1})u' - c_{1} ^{2} +  8 c_{1} c \right),
\end{equation}
and
\begin{equation} \label{EM-phantomlike-teta1}
\kappa_{6}^{2} T^{\theta} _{ \theta}  = \frac{1}{4} \left( 6(u'^{2} - e^{-2 c r}\dot{u}^{2}) - \frac{4}{3}(2 c - c_{1})u' + 16 c^{2} - c_{1} ^{2}  \right).
\end{equation}

The component $\kappa_{6}^{2} T_{r t}$ must be equal to $G_{r t}$. This energy-momentum component, as will be seen, does not influence in the general results here since we will consider only time averaged features of the above quantities. This will be better explained later in section ($4$).

Finally, the bulk cosmological constant will assume the relation
\begin{equation} \label{cons.cosm}
\Lambda _{6} = -\frac{1}{4} (24 c^{2} + c_{1} ^{2} ).
\end{equation}

This will imply in $\Lambda _{6} < 0$ which allows us to obtain relations between $c$, $c_{1}$, and $\vert \Lambda _{6} \vert$. Namely,
\begin{equation} \label{c-inequ}
c_{1} =  \pm \sqrt{4 \vert \Lambda _{6} \vert - 24 c^{2}} ,
\end{equation}
where
\begin{equation} \label{delta}
c^{2}  \leq \frac{1}{6} \vert \Lambda _{6} \vert.
\end{equation}

It may be useful to highlight that for the configuration  $A(r) = 2 c r$ and $\hspace{10pt} B(r) = c_{1} r$, the metric (\ref{gen-metric}) will assume the simpler form
\begin{equation}
\label{metric1}
ds^{2}= e^{2 c r}\left( dt^{2} - e^{u}dx^{2} - e^{u}dy^{2} - e^{-3u}dz^{2} \right) - dr^{2} - R_{0} ^{2}e^{c_{1} r + u}d\theta^{2}.
\end{equation}

Here $c$ and $c_{1}\in R$ are real constants. The range of the variables $r$ and $\theta$ are $0\leq r < \infty$ and $0\leq \theta < 2\pi$, respectively. The function $u = u(r,t)$ depends only on the variables $r$ and $t$. The compact dimension $\theta$, different of the string-like defect model, lives on the brane, i.e., $\theta$ is a brane coordinate for $r=0$. This particular feature is called hybrid compactification \cite{hybrid}.

The metric ansatz $(\ref{metric1})$ is a combination of the 6D warped braneworld model through the $e^{2 c r}$ and $e^{c_{1} r}$ terms (particularly this is similar to the global string-like defect) \cite{Oda:2000zc,Oda2000a,Gregory:1999gv,Koley2007} plus an anisotropic (r, t)-dependent warping of the brane coordinates, x, y, and z, through the terms $e^{u(t,r)}$ and $e^{ -3 u(t,r)}$. This may be seen as a six dimensional generalization of the 5D standing wave braneworld model \cite{Merab2012,Merab2011a,Merab2011b,GOGBERASHVILI1,Gogberashvili:2010yf} and a 6D generalization of the six dimensional standing wave braneworld \cite{Luis2012c,Midodashvili2012}. Still we can see our model as an extension of the global string-like defect \cite{Oda:2000zc,Oda2000a}. Therefore, for $u=0$, the metric (\ref{metric1}) is the same of the thin global string-like defects \cite{Oda:2000zc,Oda2000a}. As will be seen there are more than one point where $u=0$. In these points the geometry is the so-called $AdS$ island.

In addition, we can consider the exponential of the function $u(r,t)$  as a correction of the string-like models resulting in an anisotropic, time-dependent braneworld.

\section{Standing waves solution}
\label{solution}

In order to obtain a standing wave solution we rewrite here the differential equation for the $u(r,t)$ function (\ref{Einstein-simplif1}) as
\begin{equation}
 \label{Einstein-simplif2}
 e^{- 2 c r}\ddot{u}(r,t) -  a u^{'}(r,t) - u^{''}(r,t) = 0,
\end{equation}
where prime and dots mean differentiation with respect to $r$ and $t$, respectively, and $a = \frac{1}{6}(22 c + 4 c_{1})$. In order to solve equation $(\ref {Einstein-simplif2})$ we proceed as in  Ref. \cite{GOGBERASHVILI1} by choosing $u(r,t) = \sin(\omega t) \rho(r)$. The general solution to the  equation for the variable $\rho(r)$ is given by
\begin{equation} \label{bessel1sol_r}
\rho(r) = D_{1} e^{-\frac{a}{2} r} J_{-\frac{ a}{2 c}} (\frac{\omega}{c} e^{- c r}) + D_{2} e^{-\frac{a}{2} r} J_{\frac{ a}{2 c}} (\frac{\omega}{c} e^{- c r}),
\end{equation}
where $D_{1} = C_{1} (\omega  /2 c)^{a/2 c} \Gamma(1 - a/2 c)$ and $D_{2} = C_{2} (\omega  /2 c)^{a/2 c} \Gamma(1 + a/2 c)$.  $C_{1}$ and $C_{2}$ are integration constants. $J_{-\frac{ a}{2 c}}$ and $J_{\frac{ a}{2 c}}$ are the first kind Bessel functions of orders $-\frac{ a}{2 c}$ and $\frac{ a}{2 c}$, respectively, and  $\Gamma$ represents the Gamma function. Now that we found the solution  (\ref{bessel1sol_r}) we have the so called  standing waves solution which generalizes the 5D work \cite{GOGBERASHVILI1}, and the 6D works \cite{Luis2012c,Midodashvili2012}.  Depending on the values of  $c$ and $a$ one can obtain solutions similar to that in six dimensions. If one has $a = 5 c$ and $D_{1} = 0$ the solution will depends on the Bessel function $J_{\frac{ 5}{2 }}$ which is the case in the works in six dimensions. So these present solutions are more general then those obtained in the works cited above.

Some features of the function $u(r,t)$ can be derived from the above solution. The first one is the fact that both functions $J_{-\frac{ a}{2 c}}$ and $J_{\frac{ a}{2 c}}$ are regular at the origin and at infinity ($r \rightarrow \infty$), given the possibility to maintain the general solution  (\ref{bessel1sol_r}). Depending on the relation between $\omega$, $c$ and $a$ the functions $J_{-\frac{ a}{2 c}}$ and $J_{\frac{ a}{2 c}}$ converge for both $c>0$ or $c<0$ enabling  solutions with decreasing and increasing warp factor. Furthermore, we require that the function $u$ is zero on the brane, i.e., at $r = 0$ \cite{GOGBERASHVILI1}. This assumption may be expressed by
\begin{equation}
\label{bound-condic.}
\frac{\omega}{c} =  X_{\pm \frac{ a}{2 c}, n},
\end{equation}
where $X_{\pm \frac{ a}{2 c}, n}$ represents the n-th zero of $J_{-\frac{ a}{2 c}}$ or $J_{\frac{ a}{2 c}}$ depending if we do $C_{1}$ or $C_{2}$ equal to zero in $(\ref{bessel1sol_r})$. The boundary condition (\ref{bound-condic.}) quantizes the $\omega$ frequency.

By this consideration the $u$ function will assume the value zero in some specific $r$ values, namely $r_{m}$. For these $r_{m}$ values our model may be identified with other 6D braneworld models \cite{Gherghetta:2000qi,Giovannini:2001hh,deCarlos:2003nq,Ponton:2000gi,Oda:2000zc,Oda2000a,Koley2007,Gregory:1999gv} as one can  see in the metric (\ref{metric1}). For $c > 0$ the convergence of the function (\ref{bessel1sol_r}) for $C_{1} = 0$ or  $C_{2} = 0$, will depends essentially on the value of the ratio $\omega /c$. The quantity of zeros, or AdS island, will depend on the value of $c$ and mainly on the value of this ratio. For the case discussed here we have a finite number of zero. For $c < 0$ (with either $C_{1} = 0$ or $C_{2} = 0)$ the solution will present infinite zeros.

Once we know $u$ we may obtain the components of the energy-momentum tensor. This will be done for the cases where $a$ and $c$ have the same sign and for the case where they have opposite sign.

\subsection{Case A: the same sign for a and c}

In this case we will choose $a = 4 c$ which will imply in $c_{1} = \frac{c}{2}$. Here we will consider only the time average of the energy-momentum tensor components. This option will be better explained in the section about field localization. In the case $a = 4 c$ and $D_{1} = 0$, the solution (\ref{bessel1sol_r}) will depends on $J_{2}$, so our energy-momentum components will be done in terms of this function.

\begin{figure}[htb]
 \centering 
 \fbox{\includegraphics[width=5.5cm,height=3.5cm]{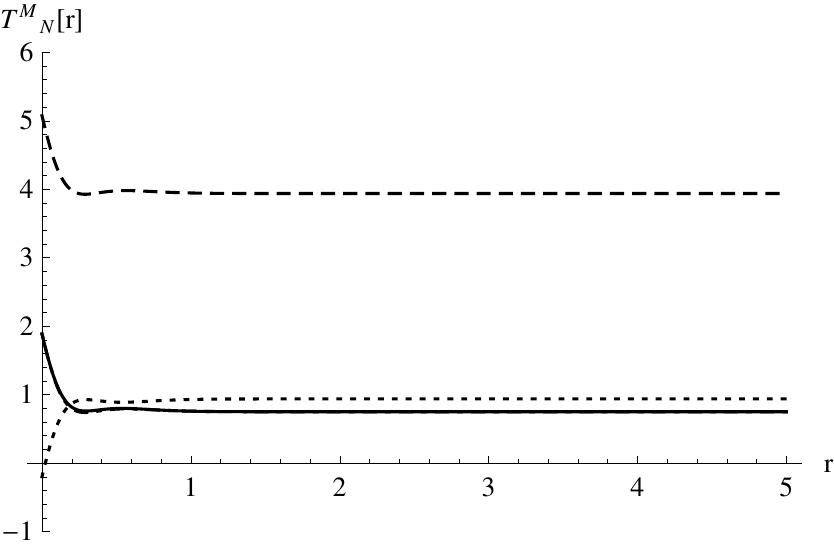}}
 \caption[Figure 1:]{$\langle T_{N} ^{M} \rangle$ profile. }
\label{fig:1}
 \end{figure}

In the figures below we plot the quantities $\langle T^{x} _{x} \rangle = \langle T^{y} _{y} \rangle = \langle T^{z} _{z} \rangle $, $ \langle T^{t} _{t} \rangle$ , $\langle T^{r} _{r} \rangle$ and $\langle T^{\theta} _{\theta} \rangle$ for $D_{2} = \kappa_{6} = c = 1$; $\omega = 5.13$. In figure (\ref{fig:1}) the dot-dashed line represents   $\langle T^{x} _{x} \rangle = \langle T^{y} _{y} \rangle = \langle T^{z} _{z} \rangle $, the doted one represents $\langle T^{r} _{r} \rangle$, the dashed line represents $\langle T^{\theta} _{\theta} \rangle$ and finally, the filled  line represents the energy density $ \langle T^{t} _{t} \rangle$. As one can see all these quantities are positive (part of $T_{r}  ^{r}$ is negative but $\vert \langle T_{r}  ^{r} \rangle \vert < \vert \langle T_{r}  ^{r} \rangle \vert$), what is an advantage when one compares it with the other works in this context \cite{Luis2012c,Midodashvili2012,GOGBERASHVILI1}. But it is not possible to say that this is a normal matter once the dominant energy condition (DEC) is violated. However it is not an exotic source once the null (NEC), strong (SEC) and weak (WEC) energy conditions are satisfied. By following the matter classification given in \cite{VISSER} this is a not normal matter.

But the we are interested in a solution generated by normal matter. In order to have a normal matter solution it is necessary that $\rho \geq p$. In order to treat this unique possibility we have to consider an anisotropic cosmological constant. As a matter of fact, recently, a higher dimensional Randall-Sundrum \textit{toy model} was proposed by Archer and Huber \cite{ARCHER} which contains a bulk with anisotropic cosmological constant given by
\begin{center}
$ \Lambda = \left( \begin{array}{c l r}
\Lambda \eta_{\mu \nu} &  &  \\
& \Lambda_{5} &  \\
 & & \Lambda_{6}
\end{array} \right)$,

\end{center}
where $\eta_{\mu \nu}$ is the metric of the brane.

Following this procedure it is possible to find our solution in the presence of normal matter. For an anisotropic cosmological constant where its  brane part is given by $\Lambda = - \frac{1}{4} ( c_{1} ^{2} + 6 c c_{1})$, $\Lambda_{5} = - \frac{1}{4} (8 c c_{1})$, and $\Lambda_{6} = - \frac{1}{4} (16 c^{2} )$, the  components of the energy-momentum tensor (\ref{EM-phantomlike-xx1} - \ref{EM-phantomlike-teta1}) will assume the form
\begin{equation} \label{EM-phantomlike-xx3}
\kappa_{6}^{2}\langle T^{x} _{x} \rangle =  \kappa_{6}^{2} \langle T^{y} _{y} \rangle = \kappa_{6}^{2} \langle T^{z} _{z} \rangle = \kappa_{6}^{2} \langle T^{\theta} _{\theta} \rangle = \frac{1}{4}\left( 6(u'^{2} - e^{-2 c r}\dot{u}^{2}) +  24 c^{2}\right),
\end{equation}

\begin{equation} \label{EM-phantomlike-tt3}
\kappa_{6}^{2} T^{t} _{t}  = - \frac{1}{4}   \left(-6(u'^{2} + e^{-2 c r}\dot{u}^{2}) -  24 c^{2} \right),
\end{equation}
and
\begin{equation} \label{EM-phantomlike-rr3}
\kappa_{6}^{2} \langle T^{r} _{r} \rangle  =  \frac{1}{4} \left( -6(u'^{2} + e^{- 2 c r}\dot{u}^{2}) +  24 c^{2}  \right).
\end{equation}

\begin{figure}[htb]
 \centering 
 \fbox{\includegraphics[width=5.5cm, height=3.5cm]{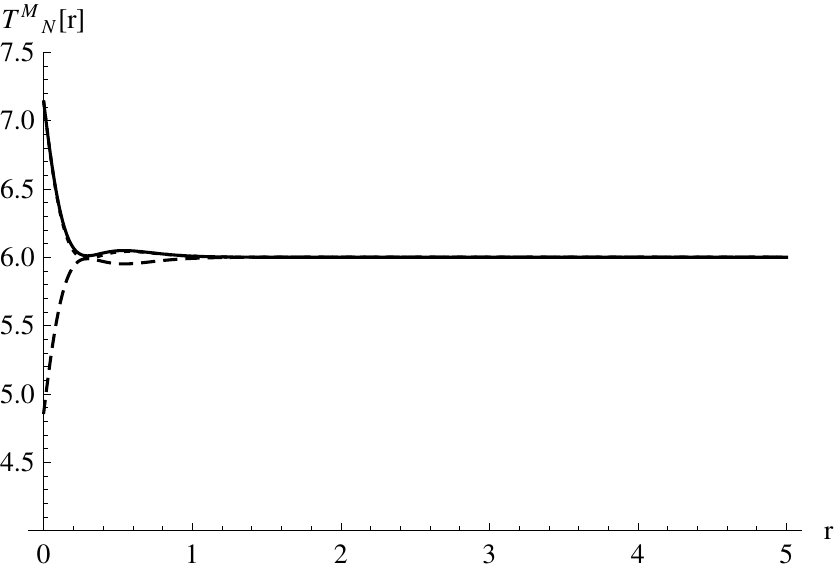}}
 \caption[Figure 1:]{$\langle T_{N} ^{M} \rangle$ profile. }

 \label{fig:2}
 \end{figure}

We plot in figure (\ref{fig:2}) these quantities as in figure (\ref{fig:1}), with the same values for the constants. The dotted line represents the spatial components, except the $r$ component which is represented by the shaded line. The filled line represents the temporal component of the energy-momentum tensor. As one can see all these quantities are positive and all the energy conditions (particularly DEC) are satisfied. This is sufficient to assure that our source is a normal matter and that our model is stable. Of course it is possible to choose the value of the cosmological constant in a different way and still keep the normal matter solution.

\subsection{Case B: a and c have opposite signs}

For this case we will consider $a = -4c$ which will give $c_{1} = - \frac{23}{2} c$. As in the other case if the cosmological constant is isotropic it is possible to find solution with all energy-momentum tensor components positive, but it would not possible to obey the dominant energy condition, as in the case above. In fact once we choose $\Lambda_{6} = - \frac{1}{4} (24 c^{2} + 6 c c_{1})$, which is positive for $a = -4c$ (meaning that the bulk is asymptotically dS), than we obtain not normal matter as in \textit{case A} above. But the principal interest consists in a source which corresponds to normal matter. So we will once again look for a solution with an anisotropic cosmological constant. There are several ways to choose the energy-momentum components and cosmological constant in order to have a solution in the presence of normal matter. Here we assume $\Lambda = - \frac{1}{4} (24 c^{2} + 6 c c_{1})$, $\Lambda_{5} = -\frac{1}{4} (24 c^{2} + 8 c c_{1} - c_{1} ^{2})$ and $\Lambda_{\theta} = -\frac{1}{4} (40 c^{2} - c_{1} ^{2})$. Since we know the relation between $c$ and $c_{1}$ it is easy to see that the components of the anisotropic cosmological constant are all positive. The time average components of the energy-momentum tensor are
\begin{equation} \label{EM-phantomlike-xx2}
\kappa_{6}^{2}\langle T^{x} _{x} \rangle =  \kappa_{6}^{2} \langle T^{y} _{y} \rangle = \kappa_{6}^{2} \langle T^{z} _{z} \rangle = \kappa_{6}^{2} \langle T^{\theta} _{\theta} \rangle = \frac{1}{4}\left( 6(u'^{2} - e^{-2 c r}\dot{u}^{2}) +  c_{1} ^{2}\right),
\end{equation}

\begin{equation} \label{EM-phantomlike-tt2}
\kappa_{6}^{2} T^{t} _{t}  = - \frac{1}{4}   \left(-6(u'^{2} + e^{-2 c r}\dot{u}^{2}) -  c_{1} ^{2} \right),
\end{equation}
and
\begin{equation} \label{EM-phantomlike-rr2}
\kappa_{6}^{2} \langle T^{r} _{r} \rangle  =  \frac{1}{4} \left( -6(u'^{2} + e^{- 2 c r}\dot{u}^{2}) +  c_{1} ^{2}  \right).
\end{equation}

\begin{figure}[htb]
 \centering 
 \fbox{\includegraphics[width=5.5cm,height=3.5cm]{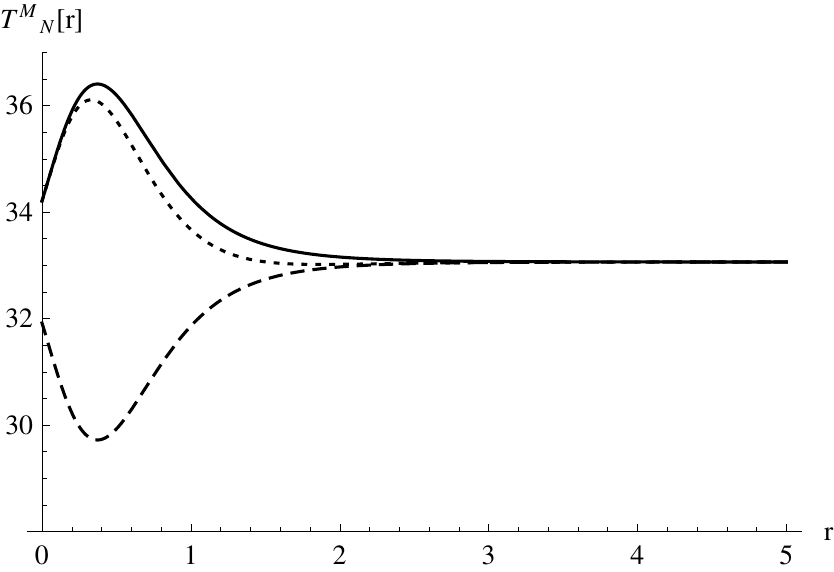}}
 \caption[Figure 1:]{$\langle T_{N} ^{M} \rangle$ profile. }

 \label{fig:3}
 \end{figure}

For $D_{2} = 0$ and $a = -4 c$ in (\ref{bessel1sol_r}) we plot the time averaged components of the energy-momentum tensor  (\ref{EM-phantomlike-tt2} - \ref{EM-phantomlike-rr2}) in figure (\ref{fig:3}). As in figure (\ref{fig:2}) the filled line represents the energy density, the dotted one gives $\kappa_{6}^{2}\langle T^{x} _{x} \rangle =  \kappa_{6}^{2} \langle T^{y} _{y} \rangle = \kappa_{6}^{2} \langle T^{z} _{z} \rangle = \kappa_{6}^{2} \langle T^{\theta} _{\theta} \rangle$ and the dashed line represents the $\langle T^{r} _{r} \rangle$ component. As one can see all these quantities are positive and $\rho \geq p$ which assures the dominant energy condition. Therefore, again we obtained a standing wave solution generated by normal matter.

On the other hand, one important feature of the braneworld models is the possibility to solve the hierarchy problem. In others standing wave braneworld works this feature was not explored. Here we are interested in showing that it is possible to readdress this solution in this context. The condition to solve the hierarchy problem in this context is that the integral below be convergent, namely
\begin{equation}
M_{4} ^{2} = 2 \pi M_{6} ^{4} \int_{0} ^{\infty} dr e^{(2c + \frac{c_{1}}{2})r}.
\end{equation}

For $c_{1} = c = 2 a$, where $a$ is a positive constant, as in the two 6D standing wave braneworld models cited above, the integral is not convergent and the hierarchy problem is not solvable. The same is valid for the first case presented in this work where $c_{1} = \frac{1}{2}c$ and $c > 0$. But in this second case where $c_{1} = - \frac{23}{2}c$ and $c$ is positive we obtain the hierarchy problem solution. So this is another advantage of the model presented here in relation to the other ones done in this same context.

After we obtain the solutions for the standing wave braneworld and after we demonstrate that one of our solutions is able to solve the hierarchy problem, we are now interest in the potential of our model in order to localize the Standard Model (SM) fields.

In the other 6D standing wave braneworld \cite{Luis2012c,Midodashvili2012} the localization issues of scalar, vector and fermion fields were already exhaustively studied. Therefore, since from our model we may obtain these other 6D standing wave braneworld, here it is sufficient to assure that the scenario presented here is convenient to localizes standard model fields. However, we will briefly present the study of localization for the scalar and fermion fields. This last one is interesting because, in five dimension, it was not possible to localize the right fermion.

\section{Scalar field localization}
\label{scalar}

This section is devoted  to the study of the localization of the bulk scalar field. We will follow again the proceedings given in Refs. \cite{Merab2011a,Merab2011b, GOGBERASHVILI1}. Then, considering the general metric  (\ref{gen-metric}) we have that $\sqrt{-g} = R_{0}e^{2A + B/2}$. Therefore, the equation for the scalar field may be written as
\begin{equation}
\label{scalar_eq_mot1}
\left[ \partial _{t} ^{2} - e^{-u}\left( \partial _{x} ^{2} + \partial _{y} ^{2}\right) - e^{3u}\partial _{z} ^{2} - \frac{e^{-u}}{R_{0} ^{2}}\partial _{\theta} ^{2}\right] \Phi = e^{-A - B/2} \left( e^{2A + B/2} \Phi ^{'}\right)^{'}.
\end{equation}

Next, we consider a solution of the form
\begin{equation}
\label{scalar_var_sep}
\Phi (x^{M})= \Psi(r, t) \chi(x, y) \zeta(z) e^{il\theta}.
\end{equation}

If one separates the variables $r$ and $t$ by making $\Psi (r,t) = e^{iEt}\bar{\rho}(r)$,  the equation for the $r$ variable will assume the form
\begin{equation}
\label{syst4}
\left( e^{2A + B/2} \bar{\rho}(r) ^{'}\right)^{'} - e^{A + B/2} G(r) \bar{\rho}(r) = 0,
\end{equation}
where
\begin{equation}
\label{Gr}
G(r) = \left( p _{x} ^{2} + p _{y} ^{2}\right)\left(e^{-u} -1\right) + p _{z} ^{2} \left(e^{3u} -1\right) + \frac{l^{2}}{R_{0} ^{2}} e^{-u}.
\end{equation}

It will be convenient to write (\ref{syst4}) as an analogue non-relativistic quantum mechanic problem. So we will assume  the change of variable  $\bar{\rho}(r) = e^{-(A + B/4)}\bar{\Psi}(r)$. With this change we will find
\begin{equation}
\label{schr-like}
\bar{\Psi}^{''}(r)  - V(r) \bar{\Psi}(r) = 0,
\end{equation}
where
\begin{equation} \label{Pot-schr}
V(r) = \frac{1}{2} (2A^{''} + \frac{B^{''}}{2}) + \frac{1}{4} (2A^{'} + \frac{B^{'}}{2})^{2} + e^{-A} G(r).
\end{equation}

From now on, we will consider $A = 2 c r$, $B = c_{1}r$ and the simplified metric (\ref{metric1}). Next, we will obtain the rt-dependent function $\Psi$, in  order to analyze the localization of the scalar field. In other words it is necessary to solve equation (\ref{schr-like}), but this will be done only for the zero mode scalar and s-wave. This case is obtained when we assume $(l=0)$ and $E = p _{x} ^{2} + p _{y} ^{2} + p _{z} ^{2}$. Additionally  it is considered that $\omega >> E$, which justifies to perform the time-averaging of $V(r)$ reducing the number of independent variables to one, namely $r$. By applying this simplification we will find the following expansion
\begin{equation}
 \label{exp-serie-solution}
\left\langle e^{bu} \right\rangle = 1 + \sum_{n = 1} ^{+ \infty} \frac{(b)^{2n}}{2^{2n}(n!)^{2}} [D_{1} e^{-\frac{a}{2} r} J_{-\frac{ a}{2 c}} (\frac{\omega}{c} e^{-c r}) + D_{2} e^{-\frac{a}{2} r} J_{\frac{ a}{2 c}} (\frac{\omega}{c} e^{- c r})]^{2n},
\end{equation}
or
\begin{equation}
\left\langle e^{bu} \right\rangle = I_{0} (b \rho(r)),
\end{equation}
where $I_{0}$ is the modified Bessel function of the first kind. It is evident from the expression above that our problem is still very complex. As can be seen from the expression (\ref{exp-serie-solution}), the approach in order to solve analytically the equation (\ref{schr-like}) is a hard work. Our strategy consists in consider simplification and asymptotic approximations for the above expression.

Let us begin the approximations by making $D_{1} = 0$ in (\ref{bessel1sol_r}). Once we do this, the $u(r,t)$ will depends on first kind Bessel function $J_{\frac{ a}{2 c}}$. The expansion (\ref{exp-serie-solution}) will be given by
\begin{equation} \label{exp-serie-solution-J}
\left\langle e^{bu} \right\rangle = 1 + \sum_{n = 1} ^{+ \infty} \frac{(bD_{2})^{2n} e^{- a n r}}{2^{2n}(n!)^{2}} [ J_{\frac{ a}{2 c}} (\frac{\omega}{c} e^{- c r})]^{2n}.
\end{equation}

It is evident that our solution is still very general since the order of the function $J$ is $a/2 c$. This give us the advantage to choose the order of the function $J$ more convenient for our interest, since our choosing is in accordance with the relations (\ref{c-inequ}) and (\ref{delta}).  If one choose  $a = 4c$  the order of $J$ will be 2, as in the  \textit{case A} discussed above. This solution is very similar to the one first proposed in five dimensions for the localization of the scalar field, with the difference that there the authors considered the Bessel function of the second kind, $Y_{2}$, rather than $J_{2}$ \cite{GOGBERASHVILI1}.

After applying what it was discussed above, let us study  (\ref{schr-like}) by considering asymptotic approximation far from and near the brane. For the first case, $r \rightarrow + \infty$, the expression $J_{\frac{a}{2c}} = J_{2}$ goes to zero ($(\omega /c)e^{-c r} \rightarrow 0$) and the relation  $(\ref{exp-serie-solution-J})$ will be approximated as $\left\langle e^{bu} \right\rangle \approx 1$. This will result in the following simpler form for equation (\ref{schr-like}), namely
\begin{equation}
 \label{schr-like-aproxJ}
\bar{\Psi}^{''}(r)  - \frac{289}{64} c^{2} \bar{\Psi}(r) = 0,
\end{equation}
whose solution is $e^{\pm \frac{17}{8} c r}$. We choose $\bar{\Psi} = e^{- \frac{17}{8} c r}$ and $c > 0$ which is convergent for all $r$ values. This solution is similar to that one found in 5D standing wave context in the case of  scalar field localization, for this same asymptotic limit assumed here \cite{Merab2011b}.

The other case to be considered here for asymptotic approximation is the case where $r \rightarrow 0$. In this case the equation (\ref{schr-like}) may be approximated as
\begin{equation}
 \label{schr-like-small-r}
\bar{\Psi}^{''}(r)  -\left( 8 d c^{2} r^{2} - 6 d c r + d^{'}\right) \bar{\Psi}(r) = 0.
\end{equation}

This equation is more general that the equivalent equation considered in five dimension \cite{Merab2011b}, once there it was considered only first order approximation. The constants $d$ and $d^{'}$ are given, respectively, by
\begin{equation}
 \label{constant1}
d = \left(\frac{D_{2} }{4}\right)^{2} \left(\frac{\omega }{c}\right)^{4} (p_{x} ^{2} + p_{y} ^{2} + 9 p_{z} ^{2}),
\end{equation}
and
\begin{equation}
\label{constant2}
d^{'} = \frac{9}{4} c^{2} + d.
\end{equation}

The solution of the equation (\ref{schr-like-small-r}) is given by
\begin{eqnarray}
\label{sol_schr-like-small-r}
\bar{\Psi} (r)  & = &  E_{1} D_{\mu} \left( -\frac{3}{2^{7/4}} \sqrt{\frac{\sqrt{d}}{c}} + 2 \left(\sqrt{c \sqrt{ 2 d}}\right) r \right)\nonumber\\
                & + &  E_{2} D_{\nu} \left( -i \frac{3}{2^{7/4}} \sqrt{\frac{\sqrt{d}}{c}} + 2 i \left(\sqrt{c \sqrt{ 2 d}}\right) r \right),
\end{eqnarray}
where $D$ is the parabolic cylinder function, and $E_{1}$ and $E_{2}$ are integration constants. We see that $E_{2}$ must be zero in order to have a real solution. The $\mu$, $\nu $ indexes are given respectively by
\begin{equation}
\label{D-index1}
\mu = -\frac{18c^{2} + 16 \sqrt{2d}c - d}{32 \sqrt{2d}c},
\end{equation}
and
\begin{equation}
\label{D-index2}
\nu = \frac{18 \sqrt{2} c^{2} - 32 \sqrt{d}c - \sqrt{2} d}{64 \sqrt{d}c}.
\end{equation}

\begin{figure}[htb] 
       \begin{minipage}[b]{0.48 \linewidth}
           \fbox{\includegraphics[width=\linewidth]{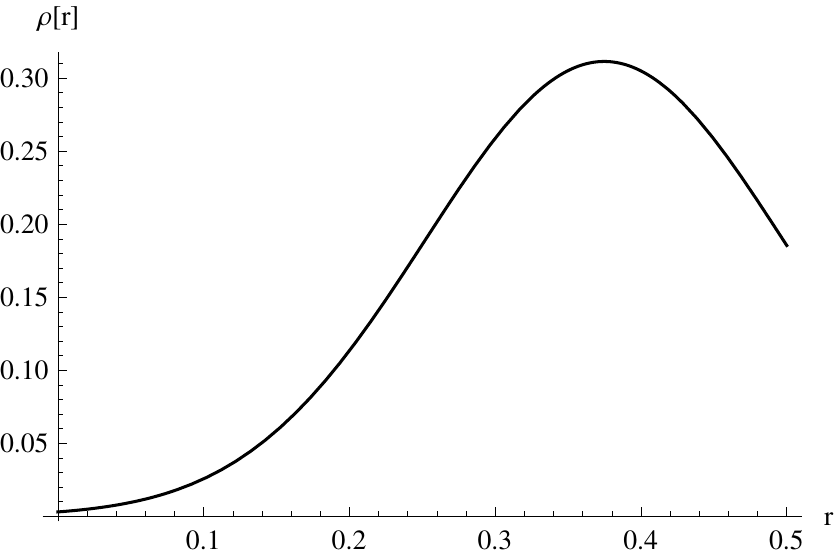}}\\
           \caption{\it $\bar{\rho}$ profile. $c = 1$}
           \label{fig:4}
       \end{minipage}\hfill
       \begin{minipage}[b]{0.48 \linewidth}
           \fbox{\includegraphics[width=\linewidth]{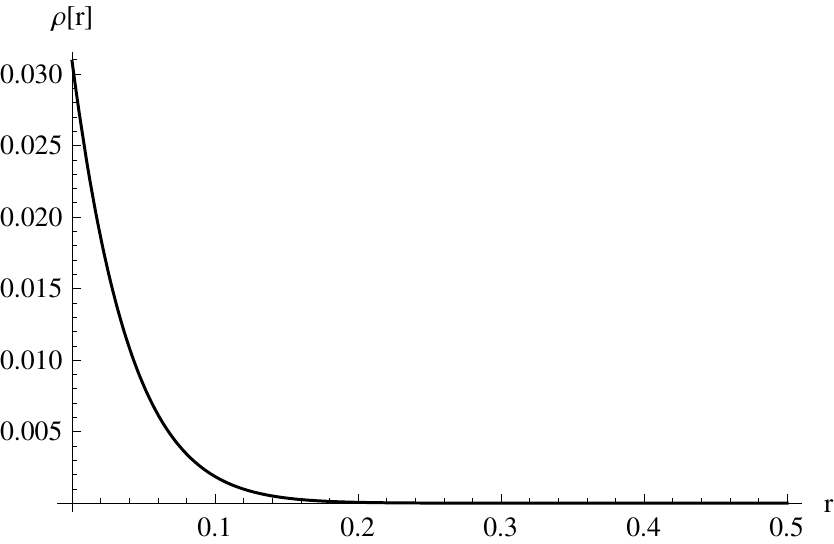}}\\
           \caption{\it $\bar{\rho}$ profile. $c = -1$}
           \label{fig:5}
       \end{minipage}
   \end{figure}

For $E_{2} = 0$ and $\omega/a = 5.13$, which corresponds to the first zero of $J_{2}$, and requiring $\mu = 0$ it  is possible to show that this solution  is convergent for either $c > 0$ or $c < 0$, as can be seen in the figures below. We see that the extra part of the scalar zero-mode wave function $\bar{\rho}(r)$ has a minimum at $r=0$, increases and then fall off, for the case $c=1$, as can be seen in figure (\ref{fig:4}). For $c=-1$ the function has a maximum at $r = 0$ and it rapidly falls off as we move away from the brane, as can be seen in figure (\ref{fig:5}). On the other hand, for $r \rightarrow \infty$, it assumes the asymptotic form  $e^{- (17/8) c r}$ which is in accordance with \cite{Merab2011b} only for $c>0$. In general, however, for other relations between $a$ and $c$, it is possible to have localization for $c$ positive or negative, i.e., for increasing or decreasing warp factor.

The results of this section show that we have  the localization of the zero-mode scalar field in the model considered in this work. This is an expected result since the study of localization of the scalar field was performed in simpler 5D and 6D models than the one considered here.  It is relevant to stress the fact  that the localization here is possible for both increasing or decreasing warp factor whereas in the thin string-like brane the localization of the zero-mode scalar field is obtained only for a decreasing warp factor, and in other standing wave braneworld this result was obtained only for the case of an increasing warp factor \cite{Oda:2000zc,Oda2000a,Luis2012c, Midodashvili2012}.

\section{Localization of spin $1/2$ fermionic zero mode }
\label{fermion}

The study of localization of zero mode spin $1/2$ fermion is interesting in this context since, in 5D standing wave braneworld, it was not possible to localize the zero mode right fermion. However in the six dimensional models cited above it was demonstrated that this field is localized. Once the model presented here is more general than that, it is natural that we find the same results here. As a matter of fact, our results, as will be seen in this section is very similar to the one found in Refs. \cite{Luis2012c, Midodashvili2012}, except that there, the Bessel function considered is $J_{\frac{5}{2}}$, and here we are using $J_{2}$. Therefore we begin by the action for the massless spin $1/2$ fermion in six dimensions which may be written as
\begin{equation} \label{fermion-action}
S = \int {d^{6} x \sqrt{-g} \bar{\Psi} i \Gamma ^{M} D_{M} \Psi}.
\end{equation}
From this action we derive the respective equation of motion, namely
\begin{equation} \label{fermion-eq-motion}
\left( \Gamma ^{\mu} D_{\mu} + \Gamma ^{r} D_{r} + \Gamma ^{\theta} D_{\theta}\right) \Psi(x^{M}) = 0.
\end{equation}
In this expression  $\Gamma ^{M}$ represents the curved gamma matrices which relate to the flat ones as
\begin{equation} \label{gam-matr}
\Gamma ^{M} = h_{\bar{M}}^{M} \gamma ^{\bar{M}},
\end{equation}
where the \textit{vielbein} $h_{\bar{M}}^{M}$ is defined as follows
\begin{equation} \label{vielbein}
g_{M N} = \eta _{\bar{M} \bar{N}}h_{M}^{\bar{M}} h_{N}^{\bar{N}}.
\end{equation}
The covariant derivative assumes the classical form
\begin{equation} \label{deri-covar}
D_{M} = \partial _{M} + \frac{1}{4} \Omega_{M} ^{\bar{M} \bar{N}} \gamma _{\bar{M}} \gamma _{\bar{N}}.
\end{equation}
The spin connection $\Omega_{M} ^{\bar{M} \bar{N}}$ in this case is defined as
\begin{eqnarray}
\label{spin-conx}
\lefteqn{\Omega_{M} ^{\bar{M} \bar{N}} = \frac{1}{2}h ^{N \bar{M}} \left(\partial _{M} h_{N} ^{\bar{N}} - \partial _{N} h_{M} ^{\bar{N}}\right) + }\nonumber\\
& & - \frac{1}{2}h ^{N \bar{N}} \left(\partial _{M} h_{N} ^{\bar{M}} - \partial _{N} h_{M} ^{\bar{M}}\right) - \frac{1}{2}h ^{P \bar{M} } h^{Q \bar{N}} h_{M} ^{\bar{R}} \left(\partial _{P} h_{Q \bar{R}} - \partial _{Q} h_{P \bar{R}}\right).
\end{eqnarray}

In order to find the relations between the curved gamma matrices and the flat gamma matrices we refer to the metric ansatz (\ref{metric1}) and use the relation (\ref{gam-matr}), which will give us the non zero results
\begin{eqnarray}
\label{gam-matr-relation}
\lefteqn{\Gamma ^{t} = e^{- c r} \gamma ^{\bar{t}}; \hspace{5 pt}  \Gamma ^{x} = e^{-c r - \frac{u}{2}} \gamma ^{\bar{x}}; \hspace{5 pt} \Gamma ^{y} = e^{-c r - \frac{u}{2}} \gamma ^{\bar{y}}; \hspace{5 pt} }\nonumber\\
& & \Gamma ^{z} = e^{-c r + \frac{3 u}{2}} \gamma ^{\bar{z}}; \hspace{5 pt} \Gamma ^{r} =  \gamma ^{\bar{r}}; \hspace{5 pt} \Gamma ^{\theta} = R_{0} ^{-1} e^{-\frac{c_{1}}{2} r - \frac{u}{2}} \gamma ^{\bar{\theta}}.
\end{eqnarray}

It is still necessary to put in evidence the non-vanishing components of the spin connection (\ref{spin-conx}), namely
\begin{eqnarray}
\label{spin-conx-nonvan}
\lefteqn{\Omega_{x} ^{\bar{t} \bar{x}} = \Omega_{y} ^{\bar{t} \bar{y}} = \frac{1}{R_{0}} \Omega_{\theta} ^{\bar{t} \bar{\theta}} = - \frac{\dot{u}}{2} e^{u/2}; \hspace{5 pt} \Omega_{z} ^{\bar{t} \bar{z}} =  \frac{3 \dot{u}}{2} e^{- 3 u/2};}\nonumber\\
& &    \Omega_{x} ^{\bar{r} \bar{x}} = \Omega_{y} ^{\bar{r} \bar{y}} = \left(c + \frac{ u^{'}}{2} \right) e^{c r +  u/2}; \hspace{5 pt}\Omega_{z} ^{\bar{r} \bar{z}} = \left(c - \frac{3 u^{'}}{2} \right) e^{c r - 3 u/2};  \nonumber\\
& & \frac{1}{R_{0}} \Omega_{\theta} ^{\bar{r} \bar{\theta}} = \left(\frac{c_{1}}{2} + \frac{ u^{'}}{2} \right) e^{\frac{c_{1}}{2} r +  u/2}; \hspace{5 pt} \Omega_{t} ^{\bar{r} \bar{t}} = c e^{c r}.
\end{eqnarray}

On the other hand, in order to solve Eq. (\ref{fermion-eq-motion}) it will be necessary to consider the case where $ \omega >> E$, as in the last section. In addition, a time average of the equation of motion must be performed. Furthermore, we assume the decomposition $\Psi(x^{A}) = \psi(x^{\mu})\rho (r) e^{il \theta}$ for the wave function. This will allow us to write the equation of motion (\ref{fermion-eq-motion}) as
\begin{equation} \label{fermion-eq-motion2}
\left[ \mathscr{D} + \gamma ^{r} \left( \frac{3c + c_{1}}{2} + \partial _{r} \right) - e^{-c r} R_{0} ^{-1} l^{2} \left\langle e^{-u/2} \right\rangle \right] \psi(x^{\mu})\rho (r) = 0,
\end{equation}
where the operator $\mathscr{D}$ is given as
\begin{equation} \label{operator}
\mathscr{D} = e^{-a r}\left[ \left( \left\langle e^{-u/2} \right\rangle - 1 \right) \left( \gamma ^{x} \partial _{x} + \gamma ^{y} \partial _{y} \right) + \left(\left\langle e^{3u/2} \right\rangle - 1\right) \gamma ^{z} \partial _{z} \right].
\end{equation}

Once again it is not possible to analytically solve the equation above and then we have to study this equation in the limits $r \rightarrow 0$ and $r \rightarrow \infty$. It is relevant to note that the operator $\mathscr{D}$ may be approximated as $\mathscr{D} \approx 0$ in this two distinct regions. This is a consequence of the fact that  $\left\langle e^{bu} \right\rangle \approx 1$ for $r \rightarrow 0$. This result it is shown schematically in figure (\ref{fig:6}) below. The constant $b$ assumes the values $b = -0.5$ and $b = 1.5$, represented by filled and dotted line respectively. As one can see from this figure the quantity $\left\langle e^{bu} \right\rangle $ is approximately one for both values of the constant $b$. Therefore, in both cases, the equation (\ref{fermion-eq-motion2}) for the s-wave $(l = 0)$, may be simplified as
\begin{equation} \label{fermion-eq-motion3}
  \left( \frac{3c + c_{1}}{2} + \partial _{r} \right)\rho (r) = 0.
\end{equation}

\begin{figure}[htb]
 \centering 
 \fbox{\includegraphics[width=5.5cm,height=3.5cm]{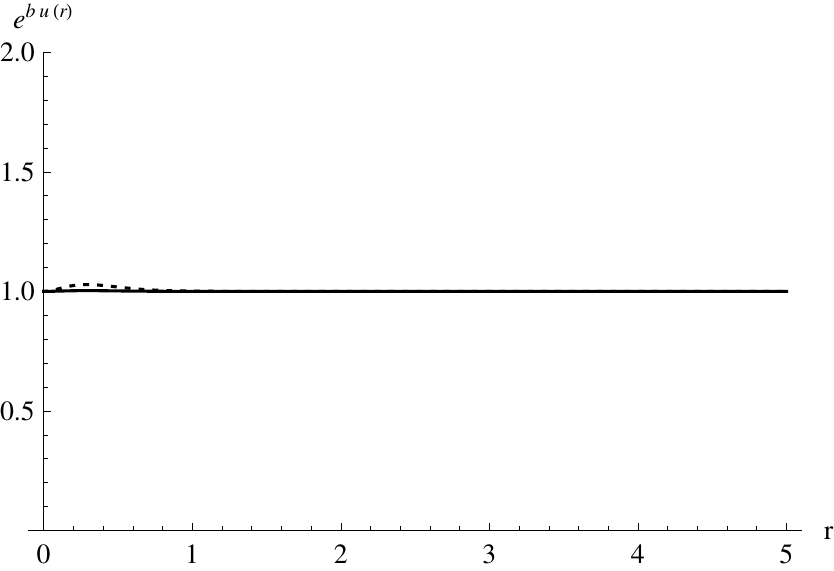}}
 \caption[Figure 1:]{$\langle e^{b u} \rangle$ profile. The filled line represents $b = - 0.5$ and the dotted line represents $b = 1.5$ }
 \label{fig:6}
 \end{figure}

It is very easy to solve this equation resulting in $\rho(r) \propto e^{-\frac{3c + c_{1}}{2} r}$.  This solution shows that for $r \rightarrow 0$ the function $\rho$ has a maximum at the origin and that it decays as $e^{-\frac{3c + c_{1}}{2} r}$ for $a > 0$ when $r \rightarrow \infty$. To show the localization, we insert this solution in the action (\ref{fermion-action}). By doing this, the resultant integral in variable $r$ will assume the form $I \propto \int_{0} ^{\infty} dr e^{-ar}$. It is evident that this integral is convergent for $a > 0$. This is sufficient  to assure that the spin $1/2$ fermion zero mode is localized in this model. Therefore, similar to the other results in six dimension, we show that the geometry is important to localize fields in contrast to 5D standing wave braneworld where is not possible to find localization for the fermion. This still shows that the model presented here is more general that the other six dimensional standing wave braneworld model and allows the localization of fields for different Bessel function.

\section{Remarks and conclusions}
\label{conclusions}

In this work, we have obtained standing gravitational waves solution for the six dimensional Einstein equation in the presence of an anisotropic brane  generated by normal matter. The compact dimension belongs to the brane and is small enough to assure that our model is realistic. Our metric ansatz is anisotropic and non-static, unlike most models considered in the braneworld literature. We find a solution for the warp factor which represents a thin brane, and in this case the bulk may be seen as a generalization of the string-like defect \cite{Oda:2000zc,Oda2000a}. Apart from having as source a physical field, our model is more general than others six dimensional standing wave braneworld recently considered in the literature \cite{Luis2012c, Midodashvili2012} since their solutions may be derived from our solution as special cases. Our metric ansatz has two different warp factors $e^{2cr}$ and $e^{2c_{1}r}$, similar to the global string-like defect, and it has also general warp factors $e^{u(r,t)}$ and $e^{-3u(r,t)}$ similar to the 6D standing wave braneworld with an exotic source. The different warp factors permit us to find a solution for the function $u$ for increasing or decreasing warp factor $e^{\pm c r}$ which is an advantage when compared with its similar model.

We find standing wave solution for the function $u(r,t)$ and we show that this solution is more general and comprises the others six dimensional standing wave braneworld solutions which have recently appeared in the literature \cite{Luis2012c, Midodashvili2012}. In fact, our solutions depend on the Bessel function $J_{\pm\frac{a}{2c}}$ which for the special case $a = \pm 5c$ coincides with the models cited above.

From the $u(r,t)$ function, it was possible to choose the type of matter that generates the brane. We have found two types of matter depending if we consider  isotropic or anisotropic cosmological constant. In the first case, for a negative cosmological constant, and $a = 4c$, we have demonstrated that the energy density and pressure components are all positive and the energy conditions NEC, WEC and SEC are satisfied, although the dominant energy condition is violated in this case. If one consider $a = -4c$, similar results are found for the matter but, in this case, the cosmological constant is positive which means that in this case the bulk is asymptotically dS, while in the first case we have an 6D AdS space-time.  It is interesting to note that an asymptotically AdS space-time has been considered in the others 5D and 6D standing wave braneworld, but a dS geometry was found for the first time here. For the case of anisotropic cosmological constant, we have considered the situation recently suggested in the literature for higher dimensional Randall-Sundrum model \cite{ARCHER}. So we consider that the cosmological constant on the brane has a fixed value $\lambda$, but  the extra components $\Lambda_{5}$ and $\Lambda_{6}$ may assume different values. This is reasonable in our case since we are dealing with an anisotropic bulk. Therefore, for $a = 4c$ all the "components" of the cosmological constant are negative and for $a = -4c$ they are positive, in line with the findings for the case of isotropic cosmological constant discussed above. In the two cases, we have obtained the components of the energy-momentum tensor and showed that all of them are positive and that all the energy conditions are satisfied. This means that we have constructed a 6D standing wave braneworld which is generated by normal matter and, therefore, stable.

An important feature of some braneworld models is their ability to solve the so-called hierarchy problem, but in the context of standing wave braneworld this problem was never treated. Here, we have demonstrated that it is possible to solve the hierarchy problem in this context. This is another advantage of the model proposed in this work over other standing wave braneworld models.

Finally, we have considered the localization of  fields in our model. Since it generalizes our previous work in this subject \cite{Luis2012c}, it is reasonable to expect that the localization of the fields which was studied there, it is also possible here. Indeed, here we have considered $a = 4c$ and we studied only the zero mode scalar and fermion fields localization and as was expected we have shown that there is a zero mode localized for both scalar and fermion fields. The solution found here for the scalar field is in accordance with the ones that were encountered in five dimensions \cite{Merab2011b} and six dimensions \cite{Luis2012c, Midodashvili2012}.

It is known that quantum effects may play important roles in braneworld models. Indeed, as an example, the mechanism of generating 4D newtonian gravity in static 3-brane \cite{dva1,dva2}. Moreover, quantum corrections in warped backgrounds may lead to gravity delocalization \cite{kaku}. Therefore, although we are dealing with normal matter, it is interesting to study how quantum fluctuation could interfere in the stability of our ansatz metric. This also will be left for a future work.

\begin{acknowledgments}
The authors thank the Funda\c{c}\~{a}o Cearense de apoio ao Desenvolvimento
Cient\'{\i}fico e Tecnol\'{o}gico (FUNCAP), the Coordena\c{c}\~{a}o de Aperfei\c{c}oamento de Pessoal de N\' ivel Superior (CAPES), and the Conselho Nacional de Desenvolvimento Cient\' ifico e Tecnol\' ogico (CNPq) for financial support.
\end{acknowledgments}


\begin{thebibliography}{99}

\bibitem{I.Antoniadis1998}
  I.~Antoniadis, N.~Arkani-Hamed, S.~Dimopoulos and G.~R.~Dvali,
  Phys.\ Lett.\ B {\bf 436}, 257 (1998)
  [hep-ph/9804398].


\bibitem{N.Arkani-Hamed1999}
  N.~Arkani-Hamed, S.~Dimopoulos and G.~R.~Dvali,
  Phys.\ Lett.\ B {\bf 429}, 263 (1998)
  [hep-ph/9803315].

\bibitem{N.Arkani-Hamed1999a}
  N.~Arkani-Hamed, S.~Dimopoulos and G.~R.~Dvali,
  Phys.\ Rev.\ D {\bf 59}, 086004 (1999)
  [hep-ph/9807344].
\bibitem{Randall1999}
  L.~Randall and R.~Sundrum,
  Phys.\ Rev.\ Lett.\  {\bf 83}, 3370 (1999)
  [arXiv:hep-ph/9905221].

\bibitem{Randall1999a}
  L.~Randall and R.~Sundrum,
  Phys.\ Rev.\ Lett.\  {\bf 83}, 4690 (1999)
  [arXiv:hep-th/9906064].

\bibitem{Oda:2000zc}
  I.~Oda,
  Phys.\ Lett.\  B {\bf 496}, 113 (2000)
  [arXiv:hep-th/0006203].


\bibitem{Oda2000a}
  I.~Oda,
  Phys.\ Rev.\ D {\bf 62}, 126009 (2000)
  [hep-th/0008012].


\bibitem{Gregory:1999gv}
  R.~Gregory,
  Phys.\ Rev.\ Lett.\  {\bf 84}, 2564 (2000)
  [arXiv:hep-th/9911015].


\cite{Chen:2000at}
\bibitem{Chen:2000at}
 J.~W.~Chen, M.~A.~Luty and E.~Ponton,
JHEP {\bf 0009}, 012 (2000)
[arXiv:hep-th/0003067].


\bibitem{Cohen:1999ia}
  A.~G.~Cohen and D.~B.~Kaplan,
  Phys.\ Lett.\  B {\bf 470}, 52 (1999)
  [arXiv:hep-th/9910132].

\bibitem{Olasagasti:2000gx}
  I.~Olasagasti and A.~Vilenkin,
  Phys.\ Rev.\ D {\bf 62}, 044014 (2000)
  [hep-th/0003300].

\bibitem{Gherghetta:2000qi}
  T.~Gherghetta and M.~E.~Shaposhnikov,
  Phys.\ Rev.\ Lett.\  {\bf 85}, 240 (2000)
  [arXiv:hep-th/0004014].

\bibitem{Ponton:2000gi}
  E.~Ponton and E.~Poppitz,
  JHEP {\bf 0102}, 042 (2001)
  [hep-th/0012033].



\bibitem{Giovannini:2001hh}
  M.~Giovannini, H.~Meyer and M.~E.~Shaposhnikov,
  Nucl.\ Phys.\ B {\bf 619}, 615 (2001)
  [hep-th/0104118].

\bibitem{Tinyakov:2001jt}
  P.~Tinyakov and K.~Zuleta,
  Phys.\ Rev.\ D {\bf 64}, 025022 (2001)
  [hep-th/0103062].


\bibitem{Kanno:2004nr}
  S.~Kanno and J.~Soda,
  JCAP {\bf 0407}, 002 (2004)
  [hep-th/0404207].

\bibitem{Vinet:2004bk}
  J.~Vinet and J.~M.~Cline,
  Phys.\ Rev.\ D {\bf 70}, 083514 (2004)
  [hep-th/0406141].

\bibitem{Cline:2003ak}
  J.~M.~Cline, J.~Descheneau, M.~Giovannini and J.~Vinet,
  JHEP {\bf 0306}, 048 (2003)
  [hep-th/0304147].

\bibitem{Papantonopoulos:2007fk}
  E.~Papantonopoulos, A.~Papazoglou and V.~Zamarias,
  Nucl.\ Phys.\ B {\bf 797}, 520 (2008)
  [arXiv:0707.1396 [hep-th]].

\bibitem{Navarro:2003vw}
  I.~Navarro,
  JCAP {\bf 0309}, 004 (2003)
  [hep-th/0302129].


\bibitem{Navarro:2004di}
  I.~Navarro and J.~Santiago,
  JHEP {\bf 0502}, 007 (2005)
  [hep-th/0411250].

\bibitem{Papantonopoulos:2005ma}
  E.~Papantonopoulos and A.~Papazoglou,
  JCAP {\bf 0507}, 004 (2005)
  [hep-th/0501112].

\bibitem{deCarlos:2003nq}
  B.~de Carlos and J.~M.~Moreno,
  JHEP {\bf 0311}, 040 (2003)
  [arXiv:hep-th/0309259].

\bibitem{Gogberashvili:2001jm}
  M.~Gogberashvili and P.~Midodashvili,
  Europhys.\ Lett.\  {\bf 61}, 308 (2003)
  [hep-th/0111132].

\bibitem{Koley2007}
  R.~Koley and S.~Kar,
  Class.\ Quant.\ Grav.\  {\bf 24}, 79 (2007)
  [hep-th/0611074].

\bibitem{Torrealba}
  R.~S.~Torrealba,
  Phys.\ Rev.\ D {\bf 82}, 024034 (2010)
  [arXiv:1003.4199 [hep-th]].

\bibitem{Silva:2011yk}
  J.~E.~G.~Silva and C.~A.~S.~Almeida,
  Phys.\ Rev.\ D {\bf 84}, 085027 (2011)
  [arXiv:1110.1597 [hep-th]].

\bibitem{Luis2012}
  L.~J.~S.~Sousa, W.~T.~Cruz and C.~A.~S.~Almeida,
  Phys.\ Lett.\ B {\bf 711}, 97 (2012)
  [arXiv:1203.5149 [hep-th]].

 \bibitem{Merab2012}
 Localization of Matter Fields in the 5D Standing Wave Braneworld, M.~Gogberashvili, arXiv:1204.2448 [hep-th].

\bibitem{Merab2011a}
  M.~Gogberashvili, P.~Midodashvili and L.~Midodashvili,
  Phys.\ Lett.\ B {\bf 707}, 169 (2012)
  [arXiv:1105.1866 [hep-th]].


\bibitem{Merab2011b}
  M.~Gogberashvili, P.~Midodashvili and L.~Midodashvili,
  Phys.\ Lett.\ B {\bf 702}, 276 (2011)
  [arXiv:1105.1701 [hep-th]].

\bibitem{Kehagias2001}
  A.~Kehagias and K.~Tamvakis,
  Phys.\ Lett.\ B {\bf 504}, 38 (2001)
  [hep-th/0010112].

\bibitem{Cruz2009}
  W.~T.~Cruz, M.~O.~Tahim and C.~A.~S.~Almeida,
  Europhys.\ Lett.\  {\bf 88}, 41001 (2009)
  [arXiv:0912.1029 [hep-th]].

\bibitem{W.T.Cruz2011}
W.~T.~Cruz, A.~R.~Gomes and C.~A.~S.~Almeida, Europhys. Lett. {\bf 96}, 31001 (2011).

\bibitem{Tahim2009}
  M.~O.~Tahim, W.~T.~Cruz and C.~A.~S.~Almeida,
  Phys.\ Rev.\ D {\bf 79}, 085022 (2009)
  [arXiv:0808.2199 [hep-th]].


\bibitem{RandjbarDaemi:2000ft}
  S.~Randjbar-Daemi and M.~E.~Shaposhnikov,
  Phys.\ Lett.\ B {\bf 491}, 329 (2000)
  [hep-th/0008087].

\bibitem{Kehagias:2004fb}
  A.~Kehagias,
  Phys.\ Lett.\  B {\bf 600}, 133 (2004)
  [arXiv:hep-th/0406025].



\bibitem{Gogberashvili:2007gg}
  M.~Gogberashvili, P.~Midodashvili and D.~Singleton,
  JHEP {\bf 0708}, 033 (2007)
  [arXiv:0706.0676 [hep-th]].

\bibitem{Duan:2006es}
  Y.~-S.~Duan, Y.~-X.~Liu and Y.~-Q.~Wang,
  Mod.\ Phys.\ Lett.\ A {\bf 21}, 2019 (2006)
  [hep-th/0602157].

\bibitem{Garriga:2004tq}
  J.~Garriga and M.~Porrati,
  JHEP {\bf 0408}, 028 (2004)
  [hep-th/0406158].

\bibitem{VazquezPoritz:2001zt}
  J.~F.~Vazquez-Poritz,
  JHEP {\bf 0209}, 001 (2002)
  [arXiv:hep-th/0111229].

\bibitem{Firouzjahi:2005qs}
  H.~Firouzjahi and S.~H.~Tye,
  JHEP {\bf 0601}, 136 (2006)
  [arXiv:hep-th/0512076].


\bibitem{Noguchi:2005ws}
  T.~Noguchi, M.~Yamaguchi and M.~Yamashita,
  Phys.\ Lett.\  B {\bf 636}, 221 (2006)
  [arXiv:hep-th/0512249].

\bibitem{Brummer:2005sh}
  F.~Brummer, A.~Hebecker and E.~Trincherini,
  Nucl.\ Phys.\  B {\bf 738}, 283 (2006)
  [arXiv:hep-th/0510113].

\bibitem{GOGBERASHVILI1}
  M.~Gogberashvili and D.~Singleton,
  Mod.\ Phys.\ Lett.\ A {\bf 25}, 2131 (2010)
  [arXiv:0904.2828 [hep-th]].

\bibitem{Gogberashvili:2010yf}
  M.~Gogberashvili, A.~Herrera-Aguilar and D.~Malagon-Morejon,
  Class.\ Quant.\ Grav.\  {\bf 29}, 025007 (2012)
  [arXiv:1012.4534 [hep-th]].

\bibitem{Caldwell et al} R. R. Caldwell, M. Kamionkowski and N. N. Weinberg, Phys. Rev. Lett. 91, 071301 (2003); S. M. Carroll, M Hoffman and M. Trodden, Phys. Rev. D68, 023509 (2003); P. Frampton,
Phys Lett. B555 139 (2003); V. Sahni and Y. Shtanov, JCAP 0311, 014 (2003).

\bibitem{Luis2012c}
A 6D standing wave Braneworld, L.~J.~S.~Sousa, J.~E.~G.~Silva and C.~A.~S.~Almeida, arXiv:1209.2727 [hep-th].

  \bibitem{VISSER}
M.~Visser, Phys. Rev. D 56, 7578 (1997).


\bibitem{Midodashvili2012} Localization of Matter Fields in the 6D Standing Wave Braneworld, P.~Midodashvili, arXiv:1211.0206v1 [hep-th].

\bibitem{ARCHER}
P.~R. ~Archer and S.~J.~Huber
JHEP 1103:018,2011.

\bibitem{hybrid} B. Carter, A. Nielsen and D. Wiltshire, JHEP \textbf{07} (2006) 34 [arxiv:0602086 [hep-th]].

\bibitem{dva1} G. Dvali, G. Gabadadze, M. Porrati, Phys. Lett. B485 (2000) 208.
\bibitem{dva2} G. Dvali, G. Gabadadze, Phys. Rev. D63 (2001) 065007.
\bibitem{kaku} Z. Kakushadze, Phys. Lett. B497 (2001) 125.



\end{thebibliography}
\end{document}